# A One-Parameter Family of Hamiltonian Structures for the KP Hierarchy and a Continuous Deformation of the Nonlinear $W_{KP}$ Algebra

José M. Figueroa-O'Farrill[1][†], Javier Mas[2][‡], and Eduardo Ramos[3][§][¶]

[1]*Physikalisches Institut der Universität Bonn, Nußallee 12, W-5300 Bonn 1, GERMANY*
[2]*Departamento de Física de Partículas Elementales, Universidad de Santiago, E-15706 Santiago de Compostela, SPAIN*
[3]*Instituut voor Theoretische Fysica, Universiteit Leuven, Celestijnenlaan 200D, B-3001 Heverlee, BELGIUM*

## ABSTRACT

The KP hierarchy is hamiltonian relative to a one-parameter family of Poisson structures obtained from a generalized Adler map in the space of formal pseudodifferential symbols with noninteger powers. The resulting W-algebra is a one-parameter deformation of $W_{KP}$ admitting a central extension for generic values of the parameter, reducing naturally to $W_n$ for special values of the parameter, and contracting to the centrally extended $W_{1+\infty}$, $W_\infty$ and further truncations. In the classical limit, all algebras in the one-parameter family are equivalent and isomorphic to $w_{KP}$. The reduction induced by setting the spin-one field to zero yields a one-parameter deformation of $\widehat{W}_\infty$ which contracts to a new nonlinear algebra of the $W_\infty$-type.

[†] e-mail: jmf@avzw03.physik.uni-bonn.de
[‡] e-mail: jamas@gaes.usc.es
[§] e-mail: fgbda06@blekul11.bitnet
[¶] Address after October 1992: Queen Mary and Westfield College, UK

§1  INTRODUCTION

The topography of W-algebras [1] in two dimensions is beginning to unfold and, among them, algebras of the $W_\infty$-type provide natural landmarks. Some of these W-algebras, which are generated by fields of integer weights $2, 3, 4, \ldots$ and possibly also 1, are expected to be universal for some infinite series of finitely generated W-algebras, in the sense [2] that all W-algebras in that series can be obtained from it as reductions. The best-known example of such a series is comprised by the $W_n$ algebras [3], of which the Virasoro algebra (corresponding to $n = 2$) is the simplest.

These algebras can be realized classically (*i.e.*, as Poisson algebras) as a certain natural reduction of the second Gel'fand–Dickey brackets [4]—a hamiltonian structure for the generalized KdV hierarchies (see [5] for a comprehensive review). These are the integrable hierarchies of isospectral deformations (of Lax type) of the one-dimensional differential operator $L = \partial^n + \sum_{j=0}^{n-1} u_j(z)\partial^j$ in terms of which, the Gel'fand–Dickey brackets have a very simple expression which we now briefly review.

Let us introduce the ring of pseudodifferential operators of the form

$$P = \sum_{j=-\infty}^{\text{finite}} p_j(z)\partial^j , \qquad (1.1)$$

with multiplication given by the generalized Leibniz rule (for $a = a(z)$)

$$\partial^p a = a\partial^p + \sum_{k=1}^{\infty} \frac{p(p-1)\cdots(p-k+1)}{k!} a^{(k)} \partial^{p-k} ; \qquad (1.2)$$

and let $P_- = \sum_{j=-\infty}^{-1} p_j \partial^j$ and $P_+ = P - P_-$ denote the projections onto the subrings of integral and differential operators respectively. Now let $X = \sum_{i=0}^{n-1} \partial^{-i-1} x_i$, and define the Adler map [6]

$$J(X) \equiv (LX)_+ L - L(XL)_+ = L(XL)_- - (LX)_- L , \qquad (1.3)$$

which sends $X$ linearly to

$$J(X) = \sum_{i,j=0}^{n-1} (J_{ij} \cdot x_j)\partial^i , \qquad (1.4)$$

for some differential operators $J_{ij}$. The second Gel'fand–Dickey bracket is then simply defined by

$$\{u_i(z),\, u_j(w)\} = -J_{ij}(z) \cdot \delta(z-w) . \qquad (1.5)$$

The constraint $u_{n-1}(z) = 0$ is second class and, upon reduction, (1.5) yields a local Poisson algebra which realizes $W_n$.

In the study of the isospectral deformations of the differential operator $L$, a crucial role is played by its $n^{\text{th}}$ root $L^{1/n} = \partial + \sum_{j=0}^{\infty} a_j \partial^{-j}$. In fact, the hierarchy can be defined starting from $L^{1/n}$ since there is a bijective correspondence between Lax flows of $L$ and of $L^{1/n}$. This prompts the definition of the KP hierarchy [7] as the isospectral deformations of a general pseudodifferential operator of the form $\Lambda = \partial + \sum_{j=0}^{\infty} a_j \partial^{-j}$. Operators like $L^{1/n}$ are obtained by imposing the constraint $\Lambda_-^n = 0$. Since the KP flows preserve this constraint, they induce isospectral deformations of $L^{1/n}$ and hence of $L$, and thus the KdV hierarchies are natural reductions of the KP hierarchy.

This fact, together with the relation between the $W_n$ algebras and the KdV hierarchies, suggests that the universal W-algebra for the $W_n$ series could be realized as a hamiltonian structure for the KP hierarchy. This reasoning led a number of authors to the construction of a new algebra—called $W_{\text{KP}}$ in the second reference of [8] and (a natural reduction thereof) $\widehat{W}_\infty$ in the third reference of [8]—by generalizing the Adler map to the space of pseudodifferential operators of the form $\Lambda$. Nevertheless, all attempts to obtain any of the $W_n$ algebras as reductions of $W_{\text{KP}}$ have failed; although as shown in [9] the classical limit of every $W_n$ can be recovered upon reduction from the classical limit of $W_{\text{KP}}$.

The possible physical relevance of $W_{\text{KP}}$ has been pointed out in [10], where a nonlinear $W_\infty$-type algebra was identified as the chiral symmetry algebra of the black hole conformal field theory based on the coset model $SL(2,R)/U(1)$. It was then conjectured that this chiral algebra is simply a quantization of $W_{\text{KP}}$. If this were so one could expect an infinite set of conserved charges to be present and eventually account for the maintenance of the quantum coherence of the black hole. An important step towards the elucidation of this conjecture was achieved in [11] via a remarkable transformation for the KP potentials in terms of only two bosons [12].

The construction of $W_{\text{KP}}$ immediately suggests how to construct an infinite number of hamiltonian structures for the KP hierarchy [13] (see also [14]). The $n^{\text{th}}$-power of the KP operator $\Lambda$ is a pseudodifferential operator $\Lambda^n = \partial^n + \sum_{j=-\infty}^{n-1} v_j \partial^j$ which contains the same information as the original KP operator and, since Lax flows of $\Lambda$ and $\Lambda^n$ correspond, can be used to describe the KP hierarchy. We can moreover define a Poisson structure by the extension of the Adler map to operators of the form $\Lambda^n$. This yields a hamiltonian structure for the KP hierarchy and a new algebra $W_{\text{KP}}^{(n)}$ which is not isomorphic to $W_{\text{KP}}$ under polynomial redefinitions of the fields and which, unlike $W_{\text{KP}}$, does admit a central extension. It is $W_{\text{KP}}^{(n)}$ which reduces naturally to $W_n$. What was proven in [9] is that the classical limit of the $W_{\text{KP}}^{(n)}$ does not depend on $n$. In other words, $W_{\text{KP}}^{(n)}$ for all $n = 1,2,3,\ldots$ is a deformation of the same classical algebra: $w_{\text{KP}}$. This prompts the study of deformations of $w_{\text{KP}}$ which may interpolate between the $W_{\text{KP}}^{(n)}$.

In this paper we shall focus on one such deformation—which we call $W_{\text{KP}}^{(q)}$. It is the Poisson structure induced by the extension of the Adler map to the space of pseudodifferential operators of the form $\partial^q + \sum_{j=1}^{\infty} b_j \partial^{q-j}$ for $q$ any complex number. Making sense out of this operator requires a bit of formalism concerning the manipulation of formal pseudodifferential symbols which shall be the focus of section 2. There we will also discuss the calculus of complex powers of pseudodifferential operators which will become instrumental in proving that $\mathsf{W}_{\mathrm{KP}}^{(q)}$ is a hamiltonian structure for the KP hierarchy.

Section 3 contains all our results which are directly concerned with integrable systems and the KP hierarchy, whereas in subsequent sections we will focus on more W-algebraic matters. Thus in section 3 we will prove that the extension of the Adler map to the space of formal symbols does indeed define a Poisson structure and show that the KP flows are hamiltonian relative to it. We also discuss the reductions to the KdV hierarchies as well as the bihamiltonian structure.

In section 4 we start the analysis of $\mathsf{W}_{\mathrm{KP}}^{(q)}$ as a W-algebra. We compute the algebra explicitly and we show that a natural reduction yields a one-parameter deformation $\widehat{\mathsf{W}}_{\infty}^{(q)}$ of $\widehat{\mathsf{W}}_{\infty}$. We write down the Virasoro subalgebra and investigate how the generators transform under it. We also investigate whether the deformation parameter $q$ is essential.

In section 5 we discuss how to recover other W-algebras of the $\mathsf{W}_{\infty}$-type as contractions and/or reductions of $\mathsf{W}_{\mathrm{KP}}^{(q)}$. In particular, we will show that the full structure (*i.e.*, with central extension in all spin sectors) of $\mathsf{W}_{1+\infty}$ arises as a suitable contraction of $\mathsf{W}_{\mathrm{KP}}^{(q)}$ as $q \to 0$. Moreover the algebra appears in a basis in which the truncation to $\mathsf{W}_{\infty}$ is manifest. The similar contraction of $\widehat{\mathsf{W}}_{\infty}^{(q)}$ yields a new genuinely nonlinear algebra $\mathsf{W}_{\infty}^{\#}$. Furthermore, contracting $\widehat{\mathsf{W}}_{\infty}^{(q)}$ as $q \to 1$ yields the full structure of $\mathsf{W}_{\infty}$. This provides a conclusive link between the full structure of $\mathsf{W}_{\infty}$ and algebraic structures associated to the Gel'fand–Dickey brackets. Generalizing one finds that for $N > 1$, the contraction as $q \to N$ recovers the centrally extended $\mathsf{W}_{\infty-N}$, a further truncation of $\mathsf{W}_{1+\infty}$.

In section 6 we discuss the classical limit of $\mathsf{W}_{\mathrm{KP}}^{(q)}$ and we show that it is independent of $q$ in the sense that the dependence of $q$ can be reabsorbed by a change of basis. Therefore all classical W-algebras in the one-parameter family are isomorphic to the algebra $\mathsf{w}_{\mathrm{KP}}$ defined in [**9**].

Finally we close the paper, in section 7, with a summary of our results and some concluding remarks on the emerging landscape of $\mathsf{W}_{\infty}$-type algebras.

§2  PSEUDODIFFERENTIAL SYMBOLS AND THEIR COMPLEX POWERS

For most practical purposes one can work with pseudodifferential operators as the ring of formal Laurent series in $\partial^{-1}$ with multiplication law given by (1.2). However, for two applications that we have in mind (namely, complex powers and the classical limit), it is convenient to work instead with the space of pseudodifferential symbols. In this section we will define pseudodifferential symbols and discuss their complex powers, postponing the discussion of the classical limit until we need it.

The Ring of Pseudodifferential Symbols

To every pseudodifferential operator $P$ we associate its symbol—a formal Laurent series—as follows. We first write $P$ with all $\partial$'s to the right: $P = \sum_{i \leq N} p_i(z) \partial^i$. (Each $P$ has a unique expression of this form.) Its symbol is then the formal Laurent series in $\xi^{-1}$ given by

$$P(z, \xi) = \sum_{i \leq N} p_i(z) \xi^i . \tag{2.1}$$

Symbols have a commutative multiplication given by multiplying the Laurent series; but one can define a composition law $\circ$ which recovers the multiplication law (1.2). In other words,

$$P(z, \xi) \circ Q(z, \xi) = (PQ)(z, \xi) , \tag{2.2}$$

where $PQ$ means the usual product of pseudodifferential operators. This composition is easily shown to be given by

$$P(z, \xi) \circ Q(z, \xi) = \sum_{k \geq 0} \frac{1}{k!} \frac{\partial^k P}{\partial \xi^k} \frac{\partial^k Q}{\partial z^k} . \tag{2.3}$$

For example, $\xi \circ a = a\xi + a'$ which recovers the basic Leibniz rule: $\partial a = a\partial + a'$ and which, upon iteration, gives rise to (1.2). Since we will be working with symbols throughout this paper, we will often drop from the notation the explicit mention of $z$ and $\xi$, referring to the symbol $P(z, \xi) = \sum_{i \leq N} p_i(z) \xi^i$ simply as $P$.

Symbol composition has the advantage that it is a well-defined operation on arbitrary smooth functions of $z$ and $\xi$ and can therefore be used to give meaning to such objects as the logarithm or a noninteger power of the derivative. For example, for $a = a(z)$,

$$\log \xi \circ a = a \log \xi - \sum_{j=1}^{\infty} \frac{(-1)^j}{j} a^{(j)} \xi^{-j} , \tag{2.4}$$

which shows that the commutator (under symbol composition) with $\log \xi$, denoted by $\operatorname{ad} \log \xi$, is an outer derivation on the ring of pseudodifferential symbols. Similarly, if $q$ is any complex number, not necessarily an integer, we find

$$\xi^q \circ a = \sum_{j=0}^{\infty} \begin{bmatrix} q \\ j \end{bmatrix} a^{(j)} \xi^{q-j} , \tag{2.5}$$

where we have introduced, for $q$ any complex number, the generalized binomial coefficients

$$\begin{bmatrix} q \\ j \end{bmatrix} \equiv \frac{q(q-1) \cdots (q-j+1)}{j!} . \tag{2.6}$$

Conjugation by $\xi^q$ is therefore an outer automorphism of the ring of pseudodifferential symbols, which is the integrated version of $\operatorname{ad} \log \xi$:

$$\xi^q \circ A(z, \xi) \circ \xi^{-q} = \exp(q \operatorname{ad} \log \xi) \cdot A(z, \xi) . \tag{2.7}$$

It follows from (2.5) that (left and right) multiplication by $\xi^q$ sends pseudodifferential symbols into symbols of the form $\sum_{j\leq N} p_j(z)\xi^{q+j}$. Let us denote the set of these symbols by $\mathcal{S}_q$. It is clear that $\mathcal{S}_q$ is a bimodule over the ring of pseudodifferential symbols, which for $q \in \mathbf{Z}$ coincides with the ring itself. In fact, since $\mathcal{S}_q = \mathcal{S}_p$ for $p \equiv q \bmod \mathbf{Z}$, we will understand $\mathcal{S}_q$ from now on as implying that $q$ is reduced modulo the integers. Moreover, symbol composition induces a multiplication $\mathcal{S}_p \times \mathcal{S}_q \to \mathcal{S}_{p+q}$, where we add modulo the integers. Therefore the union $\mathcal{S} = \cup_q \mathcal{S}_q$ forms a ring graded by the cylinder group $\mathbf{C}/\mathbf{Z}$.

On $\mathcal{S}_0$ one can define a trace form as follows. Let us define the residue of a pseudodifferential symbol $P(z,\xi) = \sum_{j\leq N} p_j(z)\xi^j$ by $\operatorname{res} P(z,\xi) = p_{-1}(z)$. Then one defines the Adler trace [6] as $\operatorname{Tr} P = \int \operatorname{res} P$, where $\int$ is any linear map which annihilates derivatives. It is easy to see that $\operatorname{Tr}[P, Q] = 0$, since the residue of a commutator is a total derivative. The Adler trace can be used to define a symmetric bilinear form on pseudodifferential symbols

$$\langle A, B \rangle \equiv \operatorname{Tr} A \circ B , \tag{2.8}$$

which extends to a symmetric bilinear form on all of $\mathcal{S}$. Relative to this form the dual space to $\mathcal{S}_q$ is clearly isomorphic to $\mathcal{S}_{-q}$ and it is an easy calculation to show that for $A = a\xi^{i+q} \in \mathcal{S}_q$ and $B = b\xi^{j-q} \in \mathcal{S}_{-q}$, the residue of their commutator is still a total derivative. In fact,

$$\operatorname{res}[A, B] = \left(\begin{bmatrix} i+q \\ i+j+1 \end{bmatrix} \sum_{l=0}^{i+j}(-1)^l a^{(l)} b^{(i+j-l)}\right)' ; \tag{2.9}$$

proving that the trace form extends to all of $\mathcal{S}$.

The ring $\mathcal{S}_0$ of pseudodifferential symbols splits into the direct sum of two subrings $\mathcal{S}_0 = \mathcal{R}_+ \oplus \mathcal{R}_-$, corresponding to the differential and integral symbols respectively. This decomposition is a maximally isotropic split for the bilinear form (2.8), since $\operatorname{Tr} A_\pm \circ B_\pm = 0$, where $A_\pm$ denotes the projection of $A$ onto $\mathcal{R}_\pm$ along $\mathcal{R}_\mp$. A similar split could in principle be defined in $\mathcal{S}_q$, but the induced split $\mathcal{S} = \mathcal{S}_+ \oplus \mathcal{S}_-$ is no longer a split into subrings as can be clearly seen from (2.5), since even if $q > 0$ its composition with $a(z) \in \mathcal{R}_+$ has an integral tail. We will therefore only write $P_\pm$ for $P \in \mathcal{S}_0$ a pseudodifferential symbol.

Complex Powers of a Pseudodifferential Symbol

Let $A(z,\xi) = \xi^n + \sum_{j=1}^\infty u_j(z)\xi^{n-j}$ for $n \in \mathbf{Z}$ be a pseudodifferential symbol and let $\alpha \in \mathbf{C}$ be any complex number. The purpose of this subsection is to define $A^\alpha$ and to prove the main properties that we expect powers to obey. Complex powers of pseudodifferential operators were first defined by Seeley and our treatment follows the one in [15].

The resolvent $R_\lambda$ of $A$ is the pseudodifferential symbol defined by

$$R_\lambda \circ (A - \lambda) = 1 , \tag{2.10}$$

for $\lambda \in \mathbf{C}$. Let us rewrite $A - \lambda$ as $A - \lambda = \sum_{j \geq 0} a_{n-j}(z, \xi, \lambda)$, where $a_n(z, \xi, \lambda) = \xi^n - \lambda$ and $a_{n-j}(z, \xi, \lambda) = u_j(z)\xi^{n-j}$, for $j \geq 1$. Notice that $a_{n-j}(z, t\xi, t^n\lambda) = t^{n-j}a_{n-j}(z, \xi, \lambda)$, whence the index of $a_{n-j}$ reflects its degree of homogeneity under the above rescalings. Formula (2.10) implies that the resolvent is given by $R_\lambda = \sum_{j \geq 0} b_{-n-j}(z, \xi, \lambda)$ and that its coefficients can be solved for recursively from $a_n b_{-n} = 1$ and

$$\sum_{j+k+l=r} \frac{1}{l!} \frac{\partial^l}{\partial \xi^l} b_{-n-j} \cdot \frac{\partial^l}{\partial z^l} a_{n-k} = 0 , \qquad (2.11)$$

for all $r \geq 1$. This implies that the coefficients of the resolvent are homogeneous under rescalings $b_{-n-j}(z, t\xi, t^n\lambda) = t^{-n-j} b_{-n-j}(z, \xi, \lambda)$.

For $\operatorname{Re}\alpha < 0$ let us define
$$A_\alpha(z, \xi) = \frac{i}{2\pi} \int_\Gamma \lambda^\alpha R_\lambda d\lambda$$
$$= \frac{i}{2\pi} \sum_{j \geq 0} \int_\Gamma \lambda^\alpha b_{-n-j}(z, \xi, \lambda) d\lambda , \qquad (2.12)$$

where the contour $\Gamma = \Gamma(\xi) = \Gamma_1 + \Gamma_2 + \Gamma_3$ is specified as follows. $\Gamma_1 = \{\lambda = re^{i\theta_\xi} | \infty > r > \rho\}$, $\Gamma_2 = \{\lambda = \rho e^{i\varphi} | \theta_\xi > \varphi > \theta_\xi - 2\pi\}$, and $\Gamma_3 = \{\lambda = re^{i(\theta_\xi - 2\pi)} | \rho < r < \infty\}$, where $\frac{1}{2}|\xi|^n > \rho$ and $\theta_\xi$ are such that for the given $\xi$, the contour does not contain any poles. The contour is oriented in such a way that the circular part is traversed clockwise. The power $\lambda^\alpha$ in (2.12) is defined as $\exp(\alpha \log \lambda)$ with $\log \lambda$ the branch of the logarithm with a cut for $\arg \lambda = \theta_\xi$. Because $\operatorname{Re}\alpha < 0$, the integral converges and each term in (2.12) is well-defined. We can therefore write
$$A_\alpha = \sum_{j \geq 0} a^{(\alpha)}_{\alpha n-j}(z, \xi)$$
where
$$a^{(\alpha)}_{\alpha n-j}(z, \xi) = \frac{i}{2\pi} \int_\Gamma \lambda^\alpha b_{-n-j}(z, \xi, \lambda) d\lambda , \qquad (2.13)$$

where again the indices reflect the degree of homogeneity. Indeed, for $t > 0$,
$$a^{(\alpha)}_{\alpha n-j}(z, t\xi) = \frac{i}{2\pi} \int_\Gamma \lambda^\alpha b_{-n-j}(z, t\xi, \lambda) d\lambda$$
$$= \frac{i}{2\pi} \int_{\Gamma^t} (t^n \mu)^\alpha b_{-n-j}(z, t\xi, t^n \mu) d\mu t^n$$
$$= t^{\alpha n-j} \frac{i}{2\pi} \int_{\Gamma^t} \mu^\alpha b_{-n-j}(z, \xi, \mu) d\mu$$
$$= t^{\alpha n-j} a^{(\alpha)}_{\alpha n-j}(z, \xi) , \qquad (2.14)$$

where we have used that the contour $\Gamma^t$ is homologous to $\Gamma$. This implies that
$$A_\alpha = \sum_{j \geq 0} u^{(\alpha)}_{\alpha n-j}(z) \xi^{\alpha n-j} \in \mathcal{S}_{\alpha n} . \qquad (2.15)$$

The following result shall prove very useful in what follows. We shall refer to it as the semigroup property of $A_\alpha$.

PROPOSITION 2.16. *For* $\operatorname{Re}\alpha < 0$ *and* $\operatorname{Re}\beta < 0$, *then* $A_\alpha \circ A_\beta = A_{\alpha+\beta}$.

PROOF: Let us define a contour $\Gamma' = \Gamma'_1 + \Gamma'_2 + \Gamma'_3$ as follows. $\Gamma'_1 = \{\lambda = re^{i(\theta_\xi - \varepsilon)} | \infty > r > \frac{3}{2}\rho\}$, $\Gamma'_2 = \{\lambda = \frac{3}{2}\rho e^{i\varphi} | \theta - \varepsilon > \varphi > \theta + \varepsilon - 2\pi\}$, and $\Gamma'_3 = \{\lambda = re^{i(\theta_\xi + \varepsilon - 2\pi)} | \frac{3}{2}\rho < r < \infty\}$, where $\varepsilon$ is chosen small enough so that $\Gamma'$ can also be used to define $A_\alpha$. Again the contour is oriented so that the circular part is traversed clockwise. We then use these contours to write

$$A_\alpha \circ A_\beta = -\frac{1}{4\pi^2} \int_{\Gamma'} \int_\Gamma R_\lambda \circ R_\mu \lambda^\alpha \mu^\beta d\mu d\lambda \ . \tag{2.17}$$

We now use the identity

$$R_\lambda \circ R_\mu = \frac{1}{\lambda - \mu}(R_\lambda - R_\mu) \tag{2.18}$$

to rewrite the above product as

$$\begin{aligned} A_\alpha \circ A_\beta &= -\frac{1}{4\pi^2} \int_{\Gamma'} \int_\Gamma \frac{\lambda^\alpha \mu^\beta}{\lambda - \mu} R_\lambda d\mu d\lambda + \frac{1}{4\pi^2} \int_\Gamma \int_{\Gamma'} \frac{\lambda^\alpha \mu^\beta}{\lambda - \mu} R_\mu d\lambda d\mu \\ &= \frac{i}{2\pi} \int_{\Gamma'} \lambda^{\alpha+\beta} R_\lambda d\lambda + 0 \\ &= A_{\alpha+\beta} \ , \end{aligned}$$

where we have used the fact that since the contour $\Gamma$ is inside the contour $\Gamma'$ the second integral vanishes since the integrand is holomorphic in $\lambda$. ∎

An immediate corollary of the semigroup property is that if $k$ is a positive integer, $A_{-k} = (A^{-1})^k$. In fact, since $\lambda^{-1}$ agrees on $\Gamma_1$ and $\Gamma_3$, the integral around $\Gamma$ in the definition of $A_{-1}$ reduces to the contour integral around $\Gamma_2$ which is a closed negatively-oriented contour. Using that $A \circ R_\lambda = 1 + \lambda R_\lambda$, we find

$$A \circ A_{-1} = \frac{i}{2\pi} \oint_{\Gamma_2} \lambda^{-1} d\lambda + \frac{i}{2\pi} \oint_{\Gamma_2} R_\lambda d\lambda = 1 \ , \tag{2.19}$$

since $R_\lambda$ is holomorphic in the disk bounded by $\Gamma_2$. Therefore it follows that $A_{-1} = A^{-1}$ and applying the semigroup property, $A_{-k} = A^{-k}$.

We can finally define the complex powers. Let $\alpha \in \mathbf{C}$ be an arbitrary complex number and let $k \in \mathbf{Z}$ be such that $\operatorname{Re}\alpha < k$. Seeley's proposal is to define

$$A^\alpha \equiv A^k \circ A_{\alpha-k} \ . \tag{2.20}$$

For this to make sense it should not depend on which $k$ we choose. In fact, if $k, j \in \mathbf{Z}$ are such that $\operatorname{Re}\alpha < k$ and $\operatorname{Re}\alpha < j$ then we should show that $A^k \circ A_{\alpha-k} = A^j \circ A_{\alpha-j}$. For definiteness let us assume that $k > j$. Then $p = k - j$ is a positive integer and let $\beta = \alpha - k$. Therefore what we have to show reduces to $A_\beta = A^{-p} \circ A_{\beta+p}$ for $\operatorname{Re}(\beta+p) < 0$. But this follows immediately from the semigroup property and the fact that $A^{-p} = A_{-p}$.

Notice that if $\operatorname{Re}\alpha < 0$ then we can simply take $k = 0$ and $A^\alpha = A_\alpha$; whereas for $j \in \mathbf{Z}$, $A^j$ agrees with the usual $j^{\text{th}}$ power: simply choose $k = j + 1$ and compute:

$$A^{j+1} \circ A_{j-(j+1)} = A^{j+1} \circ A_{-1} = A^{j+1} \circ A^{-1} = A^j \ . \tag{2.21}$$

Finally we prove the general group property of complex powers.

PROPOSITION 2.22. *Let $\alpha, \beta \in \mathbf{C}$. Then $A^\alpha \circ A^\beta = A^{\alpha+\beta}$.*

PROOF: Choose $j, k$ such that $\operatorname{Re}\alpha < j$ and $\operatorname{Re}\beta < k$. Then

$$\begin{aligned}
A^\alpha \circ A^\beta &= A^j \circ A_{\alpha-j} \circ A^k \circ A_{\beta-k} \\
&= A^j \circ A^k \circ A_{\alpha-j} \circ A_{\beta-k} \\
&= A^{j+k} \circ A_{\alpha+\beta-(j+k)} \\
&= A^{\alpha+\beta} \ ,
\end{aligned}$$

where we have used the fact that $R_\lambda$ commutes with $A$ and the semigroup property. ∎

It is possible (see, *e.g.*, [15] §§9 and 10) to topologize the space of symbols in such a way that the mapping $\alpha \mapsto A^\alpha$ is holomorphic. Since it agrees with the power for $\alpha \in \mathbf{Z}$ it makes sense to consider it as the symbol for an arbitrary complex power. In particular, a standard density argument allows us to prove that all the usual properties of rational powers are obeyed by the complex powers as well. In particular, for $\alpha, \beta \in \mathbf{C}$,

$$(A^\alpha)^\beta = A^{\alpha\beta} \ . \tag{2.23}$$

It is also clear from (2.15) that taking the $\alpha^{\text{th}}$ power maps symbols of the form $\xi^p + \cdots$ to symbols of the form $\xi^{\alpha p} + \cdots$.

## §3 THE KP HIERARCHY AND ITS HAMILTONIAN STRUCTURES

In this section we will prove that the generalization of the Adler map to the ring $\mathcal{S}$ of symbols defines a Poisson structure which depends on a complex parameter $q$. We then prove that the KP hierarchy is hamiltonian relative to this Poisson structure for all $q$ but that, unless $q$ is an integer, neither the reduction down to $n$-KdV not the bihamiltonian structure seem to be present.

## The Generalized Adler Map and Some Formal Geometry

In order to define the generalized Adler map, we need to briefly introduce some formal geometry on the space $\mathcal{M}_q$ of symbols $\Lambda^{(q)}$ of the form

$$\Lambda^{(q)} = \xi^q + \sum_{j=1}^{\infty} u_j(z) \xi^{q-j} . \tag{3.1}$$

The affine space $\mathcal{M}_q$ plays the role of the manifold on which the KP flows are defined. The tangent space $\mathcal{T}_q$ to $\mathcal{M}_q$ is parametrized by the infinitesimal deformations of $\Lambda^{(q)}$, which are given by symbols of the form

$$A = \sum_{j=1}^{\infty} a_j(z) \xi^{q-j} . \tag{3.2}$$

Every $A \in \mathcal{T}_q$ of the above form defines a vector field $\partial_A$ as follows. If $F$ is any function on $\mathcal{M}_q$, then

$$\partial_A F[\Lambda^{(q)}] = \frac{d}{d\epsilon} F[\Lambda^{(q)} + \epsilon A]\bigg|_{\epsilon=0} = \int \sum_{k=1}^{\infty} a_k \frac{\delta F}{\delta u_k} . \tag{3.3}$$

One-forms are parametrized by the dual space $\mathcal{T}_q^*$ of $\mathcal{T}_q$ under the bilinear form (2.8). In other words, $\mathcal{T}_q^*$ is made up of symbols of the form

$$X = \sum_{j=1}^{\infty} \xi^{j-q-1} \circ x_j . \tag{3.4}$$

Then, for $A$ and $X$ as above,

$$\operatorname{Tr} A \circ X = \sum_{j=1}^{\infty} \int a_j x_j , \tag{3.5}$$

which is clearly nondegenerate.

If we define, as usual, the gradient $dF$ of a function $F$ by

$$\operatorname{Tr} dF \circ A = \partial_A F , \tag{3.6}$$

for all $A \in \mathcal{T}_q$, then, for $A$ as in (3.2), one easily computes

$$\partial_A F = \sum_{j=1}^{\infty} \int \frac{\delta F}{\delta u_j} a_j ; \tag{3.7}$$

whence, comparing with (3.5), yields

$$dF = \sum_{j=1}^{\infty} \xi^{j-q-1} \circ \frac{\delta F}{\delta u_j} . \tag{3.8}$$

Hence the gradient of a function is a 1-form as expected.

Given a 1-form $X$ as in (3.4) we define a vector field $J^{(q)}(X)$ as follows
$$\begin{aligned}J^{(q)}(X) &= (\Lambda^{(q)} \circ X)_+ \circ \Lambda^{(q)} - \Lambda^{(q)} \circ (X \circ \Lambda^{(q)})_+ \\ &= \Lambda^{(q)} \circ (X \circ \Lambda^{(q)})_- - (\Lambda^{(q)} \circ X)_- \circ \Lambda^{(q)} \ . \end{aligned} \quad (3.9)$$

Notice that since $\Lambda^{(q)} \circ X$ and $X \circ \Lambda^{(q)}$ belong to $\mathcal{S}_0$, it makes sense to project onto their integral and/or differential parts. It is clear that $J^{(q)}(X) \in \mathcal{T}_q$, so that it defines a tangent vector. It is moreover easy to see that if $Y$ is another 1-form,
$$\operatorname{Tr} J^{(q)}(X) \circ Y = -\operatorname{Tr} X \circ J^{(q)}(Y) \ ; \quad (3.10)$$
making $J^{(q)}$ into a skewsymmetric linear map $J^{(q)} : \mathcal{T}_q^* \to \mathcal{T}_q$. This allows us to use $J^{(q)}$ to define a bracket on the functions on $\mathcal{M}_q$ as follows:
$$\{F, G\} = \operatorname{Tr} J^{(q)}(dF) \circ dG \ . \quad (3.11)$$

Equation (3.10) implies that this bracket is antisymmetric, and it is not too difficult to show directly that it satisfies the Jacobi identity. The proof of this fact is straightforward and can be adapted from the proof of the hamiltonian property of the original Adler map for the KdV hierarchies, to be found, for example, in [**5**]. Therefore we shall restrict ourselves to sketching the proof, trusting that the interested reader will have no trouble in filling in the details with the help of the existing literature.

Hamiltonian Property of the Generalized Adler Map

In fact, it is easier to describe the proof in a more general setting than that of the ring of symbols. Let $S = \cup_q S_q$ be an associative algebra (over some ground field $k$ of zero characteristic) graded by some commutative group written additively. Then $S_0$ is a subalgebra and $S_q$ is an $S_0$-bimodule. Suppose further that $S_0$ decomposes as the vector space direct sum of two subalgebras $S_0 = R_+ \oplus R_-$. Given any element $X \in S$ we denote by $X_\pm$ its projection to $R_\pm$ along $R_\mp$. Suppose further than we have a nondegenerate trace form $\operatorname{Tr} : S_0 \to k$ inducing a symmetric bilinear form $\langle X, Y \rangle = \operatorname{Tr} XY$ which is maximally split; that is, such that the subalgebras $R_\pm$ are maximally isotropic. In other words, $\operatorname{Tr} X_\pm Y_\pm = 0$. Let us now extend $\operatorname{Tr}$ to all of $S$ by letting it be zero on $S_{q \neq 0}$. Then the bilinear form extends to all of $S$ in such a way that $S_q$ and $S_{-q}$ are nondegenerately paired.

Choose an element $L \in S_q$ and define the generalized Adler map $J : S_{-q} \to S_q$ by
$$J(X) = (LX)_+ L - L(XL)_+ = L(XL)_- - (LX)_- L \ . \quad (3.12)$$

We can view this as an infinitesimal deformation $\delta_X L = J(X)$. More geometrically, however, we can view $J(X)$ as a vector field tangent to $S_q$ at $L$ as follows. Since $S_q$ is a linear space, we can identify its tangent space with $S_q$ itself and the cotangent space with $S_{-q}$ where the dual pairing between the tangent and cotangent spaces is given by the bilinear form. Then $J$ can be interpreted as a way to assign vector fields to 1-forms and $\delta_X$ is simply the Lie derivative in the direction $J(X)$. In particular, acting on functions, $\delta_X = \partial_{J(X)}$.

The first thing one should prove is that the vector fields obtained by $J$ close under Lie bracket or, in terms of the infinitesimal deformations $\delta_X$, that they too form a closed algebra.

LEMMA 3.13. *For all $X, Y \in S_{-q}$,*

$$[\delta_X , \delta_Y] = \delta_{[X,Y]_L^*} ,$$

*where $[X , Y]_L^*$ is given, modulo the kernel of $J$, by*

$$[X , Y]_L^* = \delta_X Y + X(LY)_- - (XL)_+ Y - (X \leftrightarrow Y) . \tag{3.14}$$

PROOF: See Proposition 3.2.2 in [5]. ∎

This bracket makes $S_{-q}$ into a Lie algebra as shown by the next proposition.

PROPOSITION 3.15. *The bracket $[,]_L^*$ defined by (3.14) satisfies the Jacobi identity.*

PROOF: Let $X, Y, Z \in S_{-q}$ and define $\mathrm{Jacobi}_L(X, Y, Z) \equiv \left[X , [Y , Z]_L^*\right]_L^* + \text{cyclic}$. We will show that this is zero. For simplicity we work under the assumption that $X, Y, Z$ are $L$-independent so that we have no terms of the form $\delta_X Y$. The general case is no harder to prove and we leave this as an exercise. By definition,

$$\begin{aligned}
\left[X , [Y , Z]_L^*\right]_L^* = {} & \delta_X [Y , Z]_L^* + X(LY(LZ)_-)_- - X(L(YL) + Z)_- \\
& - (XL)_+ Y(LZ)_- + (XL)_+ (YL)_+ Z - Y(LZ)_-(LX)_- \\
& + (YL)_+ Z(LX)_- + (Y(LZ)_- L)_+ X - ((YL)_+ ZL)_+ X \\
& - (Y \leftrightarrow Z) .
\end{aligned} \tag{3.16}$$

Also by definition,

$$\delta_X [Y , Z]_L^* = Y(J(X)Z)_- - (YJ(X))_+ Z - (Y \leftrightarrow Z) , \tag{3.17}$$

which we choose to write as

$$Y(L(XL)_- Z)_- - Y((LX)_- LZ)_- - (Y(LX)_+ L)_+ Z + (YL(XL)_+)_+ Z - (Y \leftrightarrow Z) .$$

Therefore we can write

$$\begin{aligned}
\mathrm{Jacobi}_L(X, Y, Z) = {} & Y(L(XL)_- Z)_- - Y((LX)_- LZ)_- - (Y(LX)_+ L)_+ Z \\
& + (YL(XL)_+)_+ Z + X(LY(LZ)_-)_- - X(L(YL) + Z)_- \\
& - \underbrace{(XL)_+ Y(LZ)_-}_{} + (XL)_+(YL)_+ Z - Y(LZ)_-(LX)_- \\
& + \underbrace{(YL)_+ Z(LX)_-}_{} + (Y(LZ)_- L)_+ X - ((YL)_+ ZL)_+ X + \text{signed} ,
\end{aligned} \tag{3.18}$$

– 12 –

where "signed" means signed permutations, which make the underbraced terms cancel. Now we notice that

$$Y(LZ)_-(LX)_- + Y((LX)_-LZ)_- + \text{signed} = Y((LX)_-(LZ)_+)_- + \text{signed},$$
$$Y(L(XL)_-Z)_- - X(L(YL)_+Z)_- + \text{signed} = Y(LXLZ)_- + \text{signed},$$

and

$$(YL(XL)_+)_+Z + (XL)_+(YL)_+Z + \text{signed} = ((YL)_-(XL)_+)_+Z + \text{signed}.$$

Plugging these into (3.18), we find

$$\text{Jacobi}_L(X,Y,Z) = -\underbrace{Y((LX)_-(LZ)_+)_- + Y(LXLZ)_-}_{} + ((YL)_-(XL)_+)_+Z$$
$$- (Y(LX)_+L)Z + \underbrace{X(LY(LZ)_-)_-}_{} + (Y(LZ)_-L)_+X$$
$$- ((YL)_+ZL)_+X + \text{signed}, \tag{3.19}$$

where the underbraced terms cancel after taking into account the signed permutations. What is left is

$$((YL)_-(XL)_+)_+Z - ((YL)_+ZL)_+X + (Y(LZ)_-L)_+X - (Y(LX)_+L)_+Z + \text{signed}.$$

The first term can be rewritten as $((YL)_-XL)_+Z$, which combines with the second (taking into account the signed permutations) to give $(YLXL)_+Z + \text{signed}$. Similarly, after permuting, the third and fourth terms rearrange to $-(YLXL)_+Z + \text{signed}$, which precisely cancels the contribution of the first two terms. ∎

Notice that for $q \neq 0$, the usual commutator does not close in $S_{-q}$, and hence cannot be used together with $[\,,\,]_L^*$ to give $S_{-q}$ a Lie bialgebra structure. This fact underlies the apparent lack of bihamiltonian structure for $q \notin \mathbf{Z}$ on which we will comment below.

For $F : S_q \to k$ a function, its gradient $dF \in S_{-q}$, and $J(dF)$ would be the hamiltonian vector field associated to $F$. The Poisson bracket defined by $J$ is then given by $\{F, G\} = \text{Tr}\, J(dF)dG = \partial_{J(dF)}dG$. In the same way we proved (3.10), one can show that this bracket is antisymmetric. By definition, $J$ is a hamiltonian map if and only if the bracket defined above obeys the Jacobi identity. The Jacobi identity is equivalent to the vanishing of the Schouten bracket of $J$ with itself which, being a tensorial operation, means that the Jacobi identity is a condition at a point. It then follows that it is sufficient to show it for linear functions since the gradient of any function can be substituted at a point by the gradient of a linear function.

Every $X \in S_{-q}$ independent of $L$ defines uniquely a linear function $F_X \equiv \text{Tr}\, LX$ whose gradient is given by $X$.

THEOREM 3.20. *For $X, Y, Z$ independent of $L$,*

$$\{F_Z, \{F_X, F_Y\}\} + \text{cyclic} = 0.$$

PROOF: If $F_X$ and $F_Y$ are linear functions, their Poisson bracket is given by $\{F_X, F_Y\} = \operatorname{Tr} J(X)Y$. We now use the fact that for all $W, Z$,

$$\operatorname{Tr} J(W)Z = \tfrac{1}{2} \operatorname{Tr} L \left([W, Z]_L^* - \delta_W Z + \delta_Z W\right) , \qquad (3.21)$$

which follows from (3.14) by explicit computation. In our case, since $X, Y$ are independent of $L$, it follows that $\{F_X, F_Y\} = \tfrac{1}{2} \operatorname{Tr} L [X, Y]_L^*$. Notice that since $[X, Y]_L^*$ explicitly depends on $L$, linear functions don't close under Poisson bracket. If $F_Z$ is another linear function then

$$\begin{aligned}\{F_Z, \{F_X, F_Y\}\} &= \tfrac{1}{2} \partial_{J(Z)} \operatorname{Tr} L [X, Y]_L^* \\ &= \tfrac{1}{2} \operatorname{Tr} J(Z) [X, Y]_L^* + \tfrac{1}{2} \operatorname{Tr} L \delta_Z [X, Y]_L^* ,\end{aligned}$$

which using (3.21) can be rewritten as

$$= \tfrac{1}{4} \operatorname{Tr} L \left[Z, [X, Y]_L^*\right]_L^* + \tfrac{1}{4} \operatorname{Tr} L \delta_Z [X, Y]_L^* .$$

Using Proposition 3.15 and (3.17) we find

$$\{F_Z, \{F_X, F_Y\}\} + \text{cyclic} = \tfrac{1}{4} \operatorname{Tr} \left((LY)_+ J(Z)Y - (YL)_- X J(Z)\right) + \text{signed} ,$$

which we choose to write as

$$\operatorname{Tr} \Big( \underbrace{(LX)_+ L(ZL)_- Y}_{} - (LX)_+ (LZ)_- LY \\ - \underbrace{(YL)_- X(LZ)_+ L}_{} + (YL)_- XL(ZL)_+ \Big) + \text{signed} .$$

Using cyclicity of the trace and permuting, the underbraced terms are easily seen to cancel, whereas the remaining ones can be written as

$$\operatorname{Tr} \left((ZL)_+ (YL)_- XL - (LX)_+ (LZ)_- LY\right) + \text{signed} . \qquad (3.22)$$

We now make us of the following lemma:

LEMMA 3.23. *For any $A, B, C \in S_0$,*

$$\operatorname{Tr} A_+ B_- C + \text{cyclic} = \operatorname{Tr} ABC .$$

PROOF: See [5] Lemma 3.2.4. ∎

Using this we can rewrite (3.22) as

$$\operatorname{Tr} (ZLYLXL - LXLZLY) - (X \leftrightarrow Y) ,$$

which clearly vanishes by cyclicity of the trace. ∎

## A One-Parameter Family of Hamiltonian Structures for the KP Hierarchy

The KP hierarchy is the hierarchy of isospectral flows of pseudodifferential symbols of the form

$$\Lambda = \xi + \sum_{j=1}^{\infty} a_j \xi^{j-1} , \qquad (3.24)$$

which are given by

$$\frac{\partial \Lambda}{\partial t_\ell} = \left[\Lambda_+^\ell , \Lambda\right] . \qquad (3.25)$$

It is straightforward to show that these flows commute and that they are hamiltonian with respect to the Poisson structure defined by the Adler map with hamiltonians $H_\ell = \frac{1}{\ell} \operatorname{Tr} \Lambda^\ell$. Indeed it is easy to show that we can substitute $dH_\ell = \Lambda^{\ell-1}$ in the Adler map to obtain

$$J(dH_\ell) = (\Lambda \circ \Lambda^{\ell-1})_+ \circ \Lambda - \Lambda \circ (\Lambda^{\ell-1} \circ \Lambda)_+ = \left[\Lambda_+^\ell , \Lambda\right] . \qquad (3.26)$$

Since the flows commute, the corresponding hamiltonians are in involution.

Now let $q \in \mathbf{C}^\times$ be any nonzero complex number and let $\Lambda^q$ denote the symbol of the $q^{\text{th}}$-power as defined in the previous section. Then the Lax flows correspond:

PROPOSITION 3.27. *For $q \in \mathbf{C}^\times$,*

$$\frac{\partial \Lambda}{\partial t} = [\Pi , \Lambda] \Leftrightarrow \frac{\partial \Lambda^q}{\partial t} = [\Pi , \Lambda^q] .$$

PROOF: It is clearly only necessary to prove it in one direction ($\Rightarrow$) since for the other direction we can replace $\Lambda$ by $\Lambda^{1/q}$ after invoking (2.23). So let us assume that $\frac{\partial \Lambda}{\partial t} = [\Pi , \Lambda]$. Then it is clear that for $k$ a positive integer, $\frac{\partial \Lambda^k}{\partial t} = [\Pi , \Lambda^k]$. Now let $k$ be a positive integer such that $\operatorname{Re} q < k$. Then $\Lambda_{q-k}$ is given by (2.12). Therefore its $t$-derivative is given by

$$\frac{\partial \Lambda_{q-k}}{\partial t} = \frac{i}{2\pi} \int_\Gamma \lambda^{q-k} \frac{\partial R_\lambda}{\partial t} d\lambda . \qquad (3.28)$$

But taking the $t$-derivative of (2.10) we find

$$\begin{aligned}\frac{\partial R_\lambda}{\partial t} &= -R_\lambda \circ \frac{\partial \Lambda}{\partial t} \circ R_\lambda \\ &= -R_\lambda \circ [\Pi , \Lambda] \circ R_\lambda \\ &= -R_\lambda \circ [\Pi , \Lambda - \lambda] \circ R_\lambda \\ &= [\Pi , R_\lambda] .\end{aligned}$$

Therefore,

$$\frac{\partial \Lambda_{q-k}}{\partial t} = [\Pi , \Lambda_{q-k}] , \qquad (3.29)$$

which together with the derivation property of the commutator yields the desired result. ∎

In particular, for the KP flows (3.25), Proposition 3.27 implies

$$\frac{\partial \Lambda^q}{\partial t_\ell} = \left[\Lambda^\ell_+, \Lambda^q\right] , \qquad (3.30)$$

which, using the semigroup property of complex powers, becomes

$$\frac{\partial \Lambda^q}{\partial t_\ell} = J^{(q)}(\Lambda^{\ell-q}) , \qquad (3.31)$$

where the generalized Adler map is evaluated at $\Lambda^{(q)} = \Lambda^q$. To prove that the KP flows are indeed hamiltonian relative to $J^{(q)}$ we have to find functions $H_i^{(q)}$ such that we can substitute $\Lambda^{\ell-q}$ for their gradients in $J^{(q)}$. It is not surprising that the naive result is true.

PROPOSITION 3.32. *Let $q \in \mathbf{C}^\times$, $\ell \in \mathbf{N}$ and let $H_\ell^{(q)} \equiv \frac{q}{\ell} \text{Tr}(\Lambda^q)^{\ell/q}$. Then we can substitute $dH_\ell^{(q)} = (\Lambda^q)^{(\ell/q-1)}$ in $J^{(q)}$.*

PROOF: From (3.8), the gradient $dH_i^{(q)}$ is given implicitly by

$$\left.\frac{q}{\ell}\frac{d}{d\epsilon}\text{Tr}(\Lambda^q + \epsilon A)^{\ell/q}\right|_{\epsilon=0} = \text{Tr}\, dH_\ell^{(q)} A . \qquad (3.33)$$

We therefore compute the LHS of this equation and read off the gradient after comparing with the RHS. Let $k \in \mathbf{Z}$ be a positive integer such that $\text{Re}\frac{\ell}{q} < k$. Then, by (2.20), $(\Lambda^q + \epsilon A)^{\ell/q} = (\Lambda^q + \epsilon A)^k \circ (\Lambda^q + \epsilon A)_{\ell/q-k}$, whence the LHS of (3.33) becomes

$$\left.\frac{q}{\ell}\frac{d}{d\epsilon}\text{Tr}(\Lambda^q + \epsilon A)^k \circ (\Lambda^q)_{\ell/q-k}\right|_{\epsilon=0} + \left.\frac{q}{\ell}\frac{d}{d\epsilon}\text{Tr}(\Lambda^q)^k \circ (\Lambda^q + \epsilon A)_{\ell/q-k}\right|_{\epsilon=0} . \qquad (3.34)$$

The first term can easily be evaluated to give

$$\frac{q}{\ell}\text{Tr}\sum_{j=0}^{k-1}(\Lambda^q)^j \circ A \circ (\Lambda^q)^{k-1-j} \circ (\Lambda^q)_{\ell/q-k} , \qquad (3.35)$$

which, using cyclicity of the trace and the fact that for any symbol $B$ and $\text{Re}\,\alpha < 0$, $B \circ B_\alpha = B_\alpha \circ B$, can be further simplified to

$$\frac{kq}{\ell}\text{Tr}(\Lambda^q)^{\ell/q-1} \circ A . \qquad (3.36)$$

The second term in (3.34) is trickier. Using (2.12),

$$(\Lambda^q + \epsilon A)_{\ell/q-k} = \frac{i}{2\pi}\int_\Gamma \lambda^{\ell/q-k} R_\lambda(\epsilon) d\lambda , \qquad (3.37)$$

– 16 –

where $R_\lambda(\epsilon) = (\Lambda^q - \lambda + \epsilon A)^{-1}$. Taking the $\epsilon$-derivative, we find

$$\frac{d}{d\epsilon} R_\lambda(\epsilon) \bigg|_{\epsilon=0} = -R_\lambda \circ A \circ R_\lambda \,, \tag{3.38}$$

where $R_\lambda \equiv R_\lambda(0)$. Standard manipulations inside the trace allow us to write the second term of (3.34) as

$$-\frac{q}{\ell} \operatorname{Tr}(\Lambda^q)^k \circ \left( \frac{i}{2\pi} \int_\Gamma \lambda^{\ell/q-k} R_\lambda^2 d\lambda \right) \circ A \,. \tag{3.39}$$

The integral is easily evaluated after noticing that $R_\lambda^2 = \frac{d}{d\lambda} R_\lambda$ and integrating by parts—the boundary contributions vanishing since $\operatorname{Re}(\ell/q - k) < 0$. Therefore we can rewrite (3.39) as

$$\frac{q}{\ell}(\ell/q - k) \operatorname{Tr}(\Lambda^q)^k \circ (\Lambda^q)_{\ell/q-(k+1)} \circ A = \frac{q}{\ell}(\ell/q - k) \operatorname{Tr}(\Lambda^q)^{\ell/q-1} \,, \tag{3.40}$$

which together with (3.36) yields

$$\operatorname{Tr}(\Lambda^q)^{\ell/q-1} \circ A \,, \tag{3.41}$$

which proves the proposition. ∎

This shows that the generalized Adler map $J^{(q)}$ defines, for any nonzero complex number $q$, a hamiltonian structure for the KP hierarchy.

Bihamiltonian Structure

It is well known that the KP and KdV hierarchies are actually bihamiltonian: one hamiltonian structure being the one obtained from the Adler map and the other structure being given by simply deforming the Adler map. For example, if we take $(q = 1)$ $\Lambda = \xi + \sum_{j=0}^\infty a_j \xi^{-j}$ and we deform the Adler map $J^{(1)}$ by shifting $\Lambda \mapsto \Lambda + \lambda$, we find that

$$J^{(1)} \mapsto J^{(1)}_\lambda = J^{(1)} + \lambda J^{(1)}_\infty \,, \tag{3.42}$$

where $J^{(1)}_\infty$ is given by

$$J^{(1)}_\infty(X) = [\Lambda\,, X_-]_+ - [\Lambda\,, X_+]_- \,. \tag{3.43}$$

Since this shift corresponds to the change of variables $a_0 \mapsto a_0 + \lambda$ and the Jacobi identity holds for arbitrary $a_j$, it means that for all $\lambda$ the new bracket satisfies the Jacobi identity. This being a quadratic identity, it means that it will contain pieces of orders 0, 1, and 2 in $\lambda$ which must vanish separately. The $\lambda$-independent terms are the old Jacobi identity for the unperturbed Adler map and the terms in $\lambda^2$ are the Jacobi identity for the new hamiltonian structure; whereas the terms linear in $\lambda$ are simply the expression of the fact that the two structures are coordinated—one of the hallmarks of integrability.

It is easy to see that this way of deforming the Adler map to obtain a further coordinated hamiltonian structure does not work for $q \notin \mathbf{Z}$. In fact, suppose that we shift $\Lambda^{(q)} \to \Lambda^{(q)} + \lambda A$, for some $A \in \mathcal{T}_q$. Then the deformed Adler map will in general contain terms quadratic in $\lambda$. If the deformation is to yield a further coordinated hamiltonian structure, the quadratic terms have to vanish. In other words,

$$(A \circ X)_+ \circ A - A \circ (X \circ A)_+ = 0 \qquad \forall X \ . \tag{3.44}$$

It is easy to show that if $q \notin \mathbf{Z}$, then the only solution to this equation is $A = 0$. Indeed, if $A \in \mathcal{T}_q$ has leading term $a\xi^{q-N}$ for some $N \in \mathbf{N}$, choose $X = \xi^{N-q} \circ x \in \mathcal{T}_q^*$. Then $(A \circ X)_+ = (X \circ A)_+ = ax$ and (3.44) simply says that $A$ and $ax$ commute. The leading term in their commutator is given by $(q - N)a(ax)'$ which has to vanish for all $x$. Since $q \notin \mathbf{Z}$, this means that $a = 0$.

For $q = n \in \mathbf{Z}$ things are nicer. In fact, we can choose $A = 1$, and the new hamiltonian structure is given by

$$J_\infty^{(n)}(X) = \left[\Lambda^{(n)},\, X_-\right]_+ - \left[\Lambda^{(n)},\, X_+\right]_- \ . \tag{3.45}$$

It is easy to show [5] that these are hamiltonian structures for the KP hierarchy with the same hamiltonians.

Reductions to KdV Hierarchies

The $n$-KdV hierarchy can be obtained by imposing the constraint $\Lambda_-^n = 0$ on the KP operator $\Lambda$. It follows from Proposition 3.27 that this constraint is preserved by the KP flows. Moreover the hamiltonian structure defined by $J^{(n)}$ induces a hamiltonian structure in this subspace, since for any $X$, $J^{(n)}(X)$ is a vector tangent to the space of $\Lambda^{(n)}$'s obeying $(\Lambda^{(n)})_- = 0$. In fact, this induced hamiltonian structure is nothing but the second Gel'fand–Dickey bracket for the $n$-KdV hierarchy. It is easy to see that the analogous constraint on $\Lambda^{(q)}$ is not preserved by the KP flows for $q \notin \mathbf{Z}$. This is not at all surprising given that on $\mathcal{S}_q$, the analogous projections to $\mathcal{R}_\pm$ are not natural. It would be very interesting to see if there are other hierarchies besides the KdV ones to which the KP hierarchy could reduce naturally in its hamiltonian formulation for $q \notin \mathbf{Z}$.

## §4 The $\mathsf{W}_{\mathrm{KP}}^{(q)}$ Algebra

In this section we start the discussion of the W-algebraic results of this paper. We first compute the $\mathsf{W}_{\mathrm{KP}}^{(q)}$ algebra explicitly from the fundamental Poisson brackets of the generalized Adler map $J^{(q)}$. We then introduce the reduction consisting in setting the field of lowest weight equal to zero and in this way obtain a one-parameter deformation of $\widehat{\mathsf{W}}_\infty$. We work out the first few brackets explicitly before and after reduction and we show that there is a Virasoro subalgebra with a $q$-dependent central charge. We also investigate whether there is or not a polynomial redefinition of the fields which relates different values of $q$. Finally we restrict to $q = n \in \mathbf{N}$ and we prove that $\mathsf{W}_{\mathrm{KP}}^{(n)}$ reduces naturally to $\mathsf{W}_n$.

## The Fundamental Poisson Brackets of the Adler Map and $\mathsf{W}^{(q)}_{\mathrm{KP}}$

We find it convenient to introduce a further parameter ($\alpha$) in the expression for $\Lambda^{(q)}$:

$$\Lambda^{(q)} = \alpha \xi^q + \sum_{j \geq 1} u_j(z) \xi^{q-j} \tag{4.1}$$

and, at the same time, rescale the generalized Adler map (3.9):

$$J^{(q)}(X) = \frac{1}{\alpha} \left( (\Lambda^{(q)} \circ X)_+ \circ \Lambda^{(q)} - \Lambda^{(q)} \circ (X \circ \Lambda^{(q)})_+ \right) . \tag{4.2}$$

Naturally, this does not spoil its hamiltonian properties and will become very useful when we discuss contractions of the resulting W-algebras. Let $X = \sum_{j \geq 1} \xi^{j-q-1} \circ x_j \in \mathcal{T}_q^*$. Since $J^{(q)}(X) \in \mathcal{T}_q$ is linear in $X$, we can expand it as

$$J^{(q)}(X) = \sum_{i,j \geq 1} (J^{(q)}_{ij} \cdot x_j) \xi^{q-i} , \tag{4.3}$$

where the $J^{(q)}_{ij}$ are some differential operators. As shown, for example, in the second reference of [8], the fundamental Poisson brackets arising from $J^{(q)}$ are given up to a sign by these differential operators

$$\{u_i(z), u_j(w)\} = -J^{(q)}_{ij}(z) \cdot \delta(z-w) . \tag{4.4}$$

It is precisely in expressions like these that W-algebras are classically realized.

The computation of the $J^{(q)}_{ij}$ is straightforward and we simply reproduce the result:

$$\begin{aligned}
J^{(q)}_{ij} = &\, \alpha \sum_{l=1}^{i} \begin{bmatrix} j-q-1 \\ j+l-1 \end{bmatrix} \begin{bmatrix} q \\ i-l \end{bmatrix} \partial^{i+j-1} - \sum_{l=1}^{i} \begin{bmatrix} i-1 \\ l-1 \end{bmatrix} u_{j+l-1}(-\partial)^{i-l} \\
&+ \sum_{l=1}^{i-1} \sum_{k=1}^{i-l} \begin{bmatrix} j-q-1 \\ j+l-1 \end{bmatrix} \begin{bmatrix} q-k \\ i-k-l \end{bmatrix} u_k \partial^{i+j-k-1} \\
&+ \sum_{l=1}^{i} \sum_{k=1}^{j+l-1} \begin{bmatrix} j-q-1 \\ j+l-k-1 \end{bmatrix} \begin{bmatrix} q \\ i-l \end{bmatrix} \partial^{i+j-k-1} u_k \\
&- \frac{1}{\alpha} \sum_{l=1}^{i-1} \sum_{k=1}^{i-l} \begin{bmatrix} i-k-1 \\ l-1 \end{bmatrix} u_{j+l-1}(-\partial)^{i-l-k} u_k \\
&+ \frac{1}{\alpha} \sum_{l=1}^{i-1} \sum_{k=1}^{i-l} \sum_{m=1}^{l+j-1} \begin{bmatrix} j-q-1 \\ j+l-m-1 \end{bmatrix} \begin{bmatrix} q-k \\ i-k-l \end{bmatrix} u_k \partial^{i+j-k-m-1} u_m .
\end{aligned} \tag{4.5}$$

This is a nonlinear algebra with fields of weights 1,2,3,... which for $q = 1$ reduces to the (centerless) $\mathsf{W}_{\mathrm{KP}}$ defined in the second reference of [8]. Therefore we refer to it as $\mathsf{W}^{(q)}_{\mathrm{KP}}$.

### The $u_1(z) = 0$ Reduction and $\widehat{\mathsf{W}}_\infty^{(q)}$

In order to obtain the $\widehat{\mathsf{W}}_\infty^{(q)}$ algebra we must constraint the field $u_1$ of weight 1 to vanish. It follows from (4.5) that for $q \neq 0$ this constraint is formally second-class, since $J_{11}^{(q)} = -q\alpha\partial$ is formally invertible.[1] The reduction is effected by going to the Dirac brackets (see, *e.g.*, the second reference of [**8**]), which are simply given by

$$\Omega_{ij}^{(q)} \equiv J_{ij}^{(q)} - J_{i1}^{(q)} \cdot \left(J_{11}^{(q)}\right)^{-1} \cdot J_{1j}^{(q)} \equiv J_{ij}^{(q)} + \delta J_{ij}^{(q)} \;, \tag{4.6}$$

where the differential operators are evaluated on the constraint submanifold defined by $u_1 = 0$. From (4.5) we read off the following Poisson brackets:

$$J_{1j}^{(q)} = \alpha \begin{bmatrix} j - q - 1 \\ j \end{bmatrix} \partial^j + \sum_{k=1}^{j-1} \begin{bmatrix} j - q - 1 \\ j - k \end{bmatrix} \partial^{j-k} u_k \;, \tag{4.7}$$

whereas $J_{i1}^{(q)} = -\left(J_{1i}^{(q)}\right)^*$, with $*$ the unique anti-involution on the ring of differential operators defined by $\partial^* = -\partial$ and $a(z)^* = a(z)$. Naturally we could have read off $J_{i1}^{(q)}$ from (4.5), but the resulting expression is rather complicated and simplifies only after some algebra using identities of the generalized binomial coefficients (2.6). From this we can immediately compute the correction to the fundamental Poisson brackets coming from the constraint:

$$\begin{aligned}
\delta J_{ij}^{(q)} = & \frac{\alpha}{q}(-1)^{i-1} \begin{bmatrix} i - q - 1 \\ i \end{bmatrix} \begin{bmatrix} j - q - 1 \\ j \end{bmatrix} \partial^{i+j-1} \\
& + \frac{(-1)^{i-1}}{q} \begin{bmatrix} i - q - 1 \\ i \end{bmatrix} \sum_{k=2}^{j-1} \begin{bmatrix} j - q - 1 \\ j - k \end{bmatrix} \partial^{i+j-k-1} u_k \\
& + \frac{(-1)^j}{q} \begin{bmatrix} j - q - 1 \\ j \end{bmatrix} \sum_{k=2}^{i-1} \begin{bmatrix} i - q - 1 \\ i - k \end{bmatrix} u_k(-\partial)^{i+j-k-1} \\
& - \frac{(-1)^i}{q\alpha} \sum_{k=2}^{i-1} \sum_{l=2}^{j-1} \begin{bmatrix} i - q - 1 \\ i - k \end{bmatrix} \begin{bmatrix} j - q - 1 \\ j - l \end{bmatrix} (-1)^k u_k \partial^{i+j-k-l-1} u_l \;.
\end{aligned} \tag{4.8}$$

Notice that despite the potential nonlocality present in (4.6), the resulting bracket is local. Computing the new bracket for $u_2$ uncovers a Virasoro subalgebra:

$$\Omega_{22}^{(q)} = \frac{\alpha}{12} q(q^2 - 1)\partial^3 + u_2 \partial + \partial u_2 \;. \tag{4.9}$$

Therefore $u_2$ generates diffeomorphisms under Poisson bracket in the following way: an infinitesimal diffeomorphism with parameter $\varepsilon$ induces a variation of the $u_j$ given by

$$\delta_\varepsilon u_j = -\Omega_{j2}^{(q)} \cdot \varepsilon = \left(\Omega_{2j}^{(q)}\right)^* \cdot \varepsilon \;. \tag{4.10}$$

---

[1] For $q = 0$, $u_1$ decouples from the algebra and we simply obtain a nonlinear algebra without central extension whose linear terms reproduce $\mathsf{W}_\infty$.

Using (4.6) and (4.8) we can compute

$$\delta_\varepsilon u_j = \alpha(-1)^{j+1}\left(\frac{1+q}{2}\begin{bmatrix}j-q-1\\j\end{bmatrix} + \begin{bmatrix}j-q-1\\j+1\end{bmatrix}\right)\varepsilon^{(j+1)} - ju_j\varepsilon' - u'_j\varepsilon$$
$$+ \sum_{k=2}^{j-1}\left(\frac{1+q}{2}\begin{bmatrix}j-q-1\\j-k\end{bmatrix} + \begin{bmatrix}j-q-1\\j-k+1\end{bmatrix}\right)(-1)^{j-k+1}u_k\varepsilon^{(j-k+1)} ,$$
(4.11)

which shows that $u_j$ is a field of weight $j$ under diffeomorphisms. These fields do not transform tensorially since they have higher derivatives of the parameter $\varepsilon$ in their transformation law, but it seems reasonable to expect that—at least for generic $q$—one could redefine the $u_{j>2}$ by adding differential polynomials of fields $u_{i<j}$ of lower weight in such a way that the unwanted terms cancel. For example, if we define $\widetilde{u}_3 = u_3 + \frac{1}{2}(2-q)u'_2$, then it transforms like a tensor of weight 3:

$$\delta_\varepsilon \widetilde{u}_3 = -3\widetilde{u}_3\varepsilon' - \widetilde{u}'_3\varepsilon .$$
(4.12)

Similarly, the generator $\widetilde{u}_4$, defined to be

$$u_4 - \tfrac{1}{2}(q-3)u'_3 + \tfrac{1}{10}(q-2)(q-3)u''_2 - \frac{(5q+7)(q-2)(q-3)}{10q(q^2-1)}u_2^2 ,$$
(4.13)

transforms as a tensor of weight 4. Notice that this transformation fails when the Virasoro central charge vanishes: $q = 0, \pm 1$. In particular this means that there is no primary basis for $\widehat{\mathsf{W}}_\infty$. Nevertheless, for $q$ different from those values, we would be surprised if a primary basis would not exist; although we have not worked out a proof. Notice also that if $q = 3$, then $u_4$ is already a tensor. This is a general feature: if $q = N$, then $u_{N+1}$ is already a tensor.

In summary, the $\mathsf{W}$-algebra defined by the Dirac brackets $\Omega_{ij}^{(q)}$ for $i, j \geq 2$ defines a classical realization of $\widehat{\mathsf{W}}_\infty^{(q)}$—a one-parameter deformation of $\widehat{\mathsf{W}}_\infty$, which corresponds to $q = 1$ [**8**].

### Are all $\mathsf{W}_{\mathrm{KP}}^{(q)}$ nonisomorphic?

As it stands $\mathsf{W}_{\mathrm{KP}}^{(q)}$ depends not just on $q$ but also on $\alpha$. However the dependence on $\alpha$ is fictitious. We can always reabsorb $\alpha$ by rescaling all the fields and also the generalized Adler map. Indeed from (4.5) it is clear that this is achieved by sending $u_j \mapsto \alpha u_j$, $J^{(q)} \mapsto \frac{1}{\alpha}J^{(q)}$. We consider two Poisson structures *equivalent* if there is a polynomial redefinition of the generators which takes one structure into a multiple of the other. Then the inessentiality of the parameter $\alpha$ can be re-expressed as saying that the Poisson structures corresponding to any two nonzero values of $\alpha$ are equivalent.

How about the parameter $q$? We have not been able to determine whether this parameter is essential. But passing to the reduced algebra $\widehat{\mathsf{W}}_\infty^{(q)}$ and looking at the central charge of the Virasoro subalgebra, it is clear that for $q = 0, \pm 1$, the algebra is not isomorphic to the other values of $q$. Working with $q$ a formal variable, we have investigated the first few brackets with the resulting conjecture.

CONJECTURE 4.14. *Let $p \neq q$. Then $\widehat{\mathsf{W}}_\infty^{(p)}$ and $\widehat{\mathsf{W}}_\infty^{(q)}$ are equivalent if and only if $p, q \notin \mathbf{Z}$.*

$\underline{q = n \in \mathbf{N} \text{ and the Reduction to } \mathsf{W}_n}$

We finish this section with a brief comment on $q = n \in \mathbf{N}$ and the algebras this yields. As remarked in the previous section when $q = n \in \mathbf{N}$ is a positive integer, it makes sense to impose the constraint $\Lambda_-^{(n)} = 0$. On this subspace, the generalized Adler map induces a Poisson structure which is nothing but the original Adler map for the $n^{\text{th}}$-order KdV hierarchy. Thus the W-algebra its fundamental Poisson brackets yield is the Gel'fand–Dickey algebra $\mathsf{GD}_n$ which, under the further constraint $u_1 = 0$, reduces to $\mathsf{W}_n$. Therefore we have a reduction $\mathsf{W}_{\text{KP}}^{(n)} \to \mathsf{GD}_n \to \mathsf{W}_n$. Alternatively we could have first reduced to $\widehat{\mathsf{W}}_\infty^{(n)}$ and then truncated to $\mathsf{W}_n$.

Moreover, it makes sense to deform the generalized Adler map to obtain the "first" Dickey–Radul bracket. This bracket gives rise to a W-algebra which can therefore be obtained as a contraction of $\mathsf{W}_{\text{KP}}^{(n)}$—namely $\lambda \to \infty$ in the analog of (3.42) for $J_\lambda^{(n)}$. The contracted hamiltonian map is given by (3.45) which can be suggestively rewritten as

$$J_\infty^{(n)}(X) = \left[\Lambda_+^{(n)}, X_-\right]_+ - \left[\Lambda_-^{(n)}, X_+\right]_-,$$

from where it follows that the resulting W-algebra breaks up as a direct sum of commuting subalgebras generated by the coefficients of the differential and integral parts of $\Lambda^{(n)}$, respectively. The W-algebra generated by the coefficients of the differential part is simply the first Gel'fand–Dickey bracket $\mathsf{GD}_n^{(1)}$, whereas the one generated by the coefficients of the integral part is again $\mathsf{W}_{1+\infty}$ without central extension.

## §5 Some Contractions and Reductions of $\mathsf{W}_{\text{KP}}^{(q)}$

The issue of Lie algebras containing a tower of infinitely increasing spin fields was launched in the context of two-dimensional conformal field theory in [16] where the algebra $w_\infty$—a contracting limit of $\mathsf{W}_n$—was seen to be isomorphic to the local algebra of two-dimensional area-preserving diffeomorphisms and to admit a nontrivial central extension only in the Virasoro sector. In [17], the $\mathsf{W}_\infty$ algebra was constructed as a deformation, the main motivation being to accommodate for nontrivial central extensions in all higher spin sectors. The method was brute force imposition of the Jacobi identity within a suitable Ansatz. Later [18] the $\mathsf{W}_{1+\infty}$ algebra was found as an extension of the previous one accommodating for an extra spin 1 field. The question of the relation of these algebras to the Poisson-bracket algebras induced by the Gel'fand–Dickey construction was partially understood in [19], where $\mathsf{W}_{1+\infty}$ was shown to arise as the first hamiltonian structure of the KP hierarchy. However, since the algebra obtained was centerless, the full connection still remained unclear. In this section we will show how to recover the full structure of $\mathsf{W}_{1+\infty}$ as a suitable contraction of $\mathsf{W}_{\text{KP}}^{(q)}$ as $q$ tends to 0 and not to $\infty$ as one would perhaps naively expect. The central extension arises by judiciously scaling the parameter $\alpha$ introduced in (4.1) and (4.2) in such a way that $\alpha$ tends to $\infty$, with $\alpha q = c$, a constant. Moreover, $\mathsf{W}_{1+\infty}$ is obtained in a basis making its truncation to $\mathsf{W}_\infty$ manifest; thus we also recover the full structure of $\mathsf{W}_\infty$. The full structure of $\mathsf{W}_\infty$ can also be recovered by a similar contraction of $\widehat{\mathsf{W}}_\infty^{(q)}$ this time as $q \to 1$. This procedure generalizes as follows: taking the limit $q \to N$ and $\alpha \to \infty$ such that $\alpha(q - N) = c$, a constant, of the (nonlocal) reduction of $\mathsf{W}_{\text{KP}}^{(q)}$ induced by setting the $N$ fields of lowest spin to zero, yields the full structure of the subalgebra $\mathsf{W}_{\infty-N}$ of $\mathsf{W}_{1+\infty}$ generated by the fields with spins greater than $N$. Finally we construct a new nonlinear algebra as a contraction of $\widehat{\mathsf{W}}_\infty^{(q)}$ as $q \to 0$ or, equivalently, as a reduction of $\mathsf{W}_{1+\infty}$.

## Centrally Extended $\mathsf{W}_{1+\infty}$ as a Contraction of $\mathsf{W}_{\text{KP}}^{(q)}$

In this subsection we will analyze the contraction $q \to 0$ of $\mathsf{W}_{\text{KP}}^{(q)}$. To this effect let us rewrite (4.1) as:

$$\Lambda^{(q)} = \alpha \xi^q + \sum_{j \geq 1} u_j(z) \xi^{q-j} \equiv \alpha \xi^q + S \tag{5.1}$$

which defines $S$. Notice that $\lim_{q \to 0} S = \Lambda_-$, where $\Lambda \equiv \Lambda^{(0)}$. We can expand (4.2) as follows:

$$\begin{aligned}J^{(q)}(X) = {}& \alpha \left[ (\xi^q \circ X)_+ \xi^q - \xi^q \circ (X \xi^q)_+ \right] \\& + (\xi^q \circ X)_+ \circ S + (S \circ X)_+ \xi^q - \xi^q \circ (X \circ S)_+ - S \circ (X \xi^q)_+ \\& + \frac{1}{\alpha} \left[ (S \circ X)_+ \circ S - S \circ (X \circ S)_+ \right] \, . \end{aligned} \tag{5.2}$$

Taking the limit we find that the terms quadratic in $S$ disappear, whereas the linear terms in $S$ contract trivially to give

$$[\Lambda_-, X]_+ - [\Lambda_-, X_+] = -[\Lambda_-, X_+]_- \, . \tag{5.3}$$

Expanding $\xi^q = 1 + q \log \xi + O(q^2)$ we find that, upon taking the limit, the $S$-independent terms yield

$$c \left[ (\log \xi \circ X)_+ + X_+ \circ \log \xi - \log \xi \circ X_+ - (X \circ \log \xi)_+ \right] \, , \tag{5.4}$$

which can be immediately rewritten as $-[c \log \xi, X_+]_-$. The limiting hamiltonian structure is therefore

$$J^{(0)}_{1+\infty}(X) = -[c \log \xi + \Lambda_-, X_+]_- \, . \tag{5.5}$$

The *c*-independent part is the standard first hamiltonian structure of the KP hierarchy [20] which was identified in [19] with a centerless $\mathsf{W}_{1+\infty}$. On the other hand, the *c*-dependent central extension is nothing but the Khesin–Kravchenko 2-cocycle of the Lie algebra of pseudodifferential operators on the circle which appeared in the context of W-algebras in [21] for the first time.

Taking the limit in the explicit expression (4.5), we find the following well-known expression for the centrally extended $\mathsf{W}_{1+\infty}$:

$$(J^{(0)}_{1+\infty})_{ij} = c(-1)^i \frac{(i-1)!(j-1)!}{(i+j-1)!} \partial^{i+j-1}$$
$$+ \sum_{l=1}^{j-1} \begin{bmatrix} j-1 \\ l \end{bmatrix} \partial^l u_{i+j-l-1} - \sum_{l=1}^{i-1} \begin{bmatrix} i-1 \\ l \end{bmatrix} u_{i+j-l-1}(-\partial)^l \, . \tag{5.6}$$

Notice that this basis for $\mathsf{W}_{1+\infty}$ makes manifest a nested sequence of subalgebras obtained by truncating the spectrum from below. For any $N$, the generators $\{u_i\}_{i>N}$ close among themselves. In particular for $N=1$ we recover the full structure of $\mathsf{W}_\infty$. The full structure of $\mathsf{W}_\infty$ also arises by first reducing to $\widehat{\mathsf{W}}^{(q)}_\infty$ and then taking the contracting limit $q \to 1$ as we will see in the next subsection. For $N \geq 2$, the resulting algebras ($\mathsf{W}_{\infty-N}$) do not contain a Virasoro subalgebra and are therefore not interesting from the point of view of extended conformal algebras. Nevertheless, as we will show at the end of the next subsection, they can be obtained by reducing $\mathsf{W}^{(q)}_{\mathrm{KP}}$ and then contracting to $q \to N$.

<u>The Full Structure of $\mathsf{W}_\infty$ as a $q \to 1$ Contraction of $\widehat{\mathsf{W}}^{(q)}_\infty$.</u>

We now investigate the contraction $q \to 1$ and $\alpha \to \infty$ in such a way that $\alpha(q-1) = c$, a constant. From (4.5) it follows that $J^{(q)}_{11}$ diverges, hence it is necessary to impose the constraint $u_1(z) = 0$. However the correction terms implied by the Dirac bracket (4.6) do not contribute since they contain a $(J^{(q)}_{11})^{-1}$ which is zero in the limit. For $i,j \geq 2$, the central term is finite in the limit, since letting $q = 1 + \varepsilon$, we see that the central term in (4.5) is $O(\varepsilon)$—the $O(1)$ terms being absent. The terms of order $\varepsilon$ all come from $l=i$ and $l=i-1$ in the sum:

$$\alpha \left( \begin{bmatrix} j-q-1 \\ i+j-1 \end{bmatrix} + q \begin{bmatrix} j-q-1 \\ i+j-2 \end{bmatrix} \right) \partial^{i+j-1} \, , \tag{5.7}$$

which in the limit yield the central terms in (5.6). The nonlinear terms in (4.5) are polynomial in $q$ and scale inversely with $\alpha$, thus vanishing in the limit. After some cosmetics, the linear terms reproduce the linear terms of (5.6). Therefore the resulting algebra is simply the subalgebra of (5.6) generated by $\{u_i\}_{i>1}$—that is, $\mathsf{W}_\infty$ with central extension.

As advertised before, this fact generalizes. If we take the limit $q \to N$ and $\alpha \to \infty$ such that $\alpha(q - N) = c$, we find that the central terms in (4.5) for $i, j \leq N$ all diverge in the limit. We must therefore reduce the algebra by setting them to zero. The resulting algebra–denoted tentatively[2] by $\widehat{\mathsf{W}}^{(q)}_{\infty-N}$) is nonlocal for all values of $q$. However in the limit, the nonlocal (as well as the nonlinear) terms all vanish and we are left—after similar manipulations to the ones described above for $N = 1$—with the subalgebra of (5.6) generated by the $\{u_i\}_{i>N}$—namely, $\mathsf{W}_{\infty-N}$ with central extension. It is possible— although we refrain from doing this here for the sake of brevity—to obtain for all $N$ an explicit expression for the limit of the reduced hamiltonian map analogous to (5.5) and featuring the Khesin-Kravchenko cocycle.

### A New Nonlinear Algebra as a Contraction of $\widehat{\mathsf{W}}^{(q)}_\infty$

Imposing the constraint $u_1(z) = 0$ on the $\mathsf{W}_{1+\infty}$ algebra we obtain $\mathsf{W}^{\#}_\infty$—a new nonlinear algebra. From the results of the preceding subsection, it follows that this path to $\mathsf{W}^{\#}_\infty$ can be summarized as

$$\mathsf{W}^{(q)}_{\mathrm{KP}} \xrightarrow{\text{contraction}} \mathsf{W}_{1+\infty} \xrightarrow{\text{reduction}} \mathsf{W}^{\#}_\infty \ . \tag{5.8}$$

Performing the operations in the reverse order we recover the same algebra, whence we can exhibit $\mathsf{W}^{\#}_\infty$ as the contraction of $\widehat{\mathsf{W}}^{(q)}_\infty$. The explicit expression of this algebra can be obtained by contracting the operators $\Omega^{(q)}_{ij}$ defined by (4.6). The contraction of $J^{(q)}_{ij}$ is simply given by (5.6), whereas the contraction of $\delta J^{(q)}_{ij}$ is given by

$$\begin{aligned}(\delta J^{(0)}_{1+\infty})_{ij} = \ & -c\frac{(-1)^i}{ij}\partial^{i+j-1} + \frac{(-1)^i}{i}\sum_{l=2}^{j-1}\begin{bmatrix}j-1\\l-1\end{bmatrix}\partial^{i+j-l-1}u_l \\ & -\frac{(-1)^j}{j}\sum_{l=2}^{i-1}\begin{bmatrix}i-1\\l-1\end{bmatrix}u_l(-\partial)^{i+j-l-1} \\ & -\frac{(-1)^i}{c}\sum_{k=2}^{i-1}\sum_{l=2}^{j-1}\begin{bmatrix}i-1\\k-1\end{bmatrix}\begin{bmatrix}j-1\\l-1\end{bmatrix}(-1)^k u_k \partial^{i+j-k-l-1} u_l \ . \end{aligned} \tag{5.9}$$

One can show that this is a genuinely nonlinear algebra in that there exists no polynomial field redefinition which linearizes it. Moreover it can be shown by inspection of the first few Poisson brackets not to be equivalent to $\widehat{\mathsf{W}}^{(q)}_\infty$ for any $q$; that is, there exists no polynomial redefintion of fields which sends $\mathsf{W}^{\#}_\infty$ to a multiple of $\widehat{\mathsf{W}}^{(q)}_\infty$ for any $q$.

---

[2] In [**22**] strong evidence was presented to suggest the existence of nonlinear local algebras denoted $\widehat{\mathsf{W}}_{\infty-N}$ with the same spectrum as the algebras we obtain here. Our choice of notation notwithstanding, we have not been able to exhibit between these algebras and the ones we obtain here any link besides the fact that they are both deformations of $\mathsf{W}_{\infty-N}$.

## §6 The Classical Limit of $\mathsf{W}_{\text{KP}}^{(q)}$

In this section we investigate the classical limit of $\mathsf{W}_{\text{KP}}^{(q)}$. The classical limit of Gel'fand–Dickey brackets (see, for example, [**9**]) is defined as the brackets induced by the Adler map in the commutative limit of the ring of pseudodifferential operators. We shall briefly sketch this, referring the reader to [**9**] for more details.

The starting point for defining the classical or commutative limit of the ring of pseudodifferential operators is the introduction in (2.3) of a formal parameter $\hbar$ as follows:

$$P \circ Q = \sum_{k \geq 0} \frac{\hbar^k}{k!} \frac{\partial^k P}{\partial \xi^k} \frac{\partial^k Q}{\partial z^k} \ , \tag{6.1}$$

interpolating from the (commutative) multiplication of symbols for $\hbar = 0$ to the (non-commutative) composition of symbols for $\hbar = 1$. The classical limit of any structure is obtained by introducing the parameter $\hbar$ via (6.1) and keeping only the lowest term in its $\hbar$ expansion. Therefore, the classical limit of $\circ$ is simply the commutative multiplication of symbols; hence the name commutative limit.

Symbols can be made into a Poisson algebra, where the Poisson bracket is defined as the classical limit of the commutator—namely,[3]

$$[\![P, Q]\!] = \lim_{\hbar \to 0} \hbar^{-1} [P, Q] \ , \tag{6.2}$$

which can be written explicitly with the help of (6.1) as

$$[\![P, Q]\!] = \frac{\partial P}{\partial \xi} \frac{\partial Q}{\partial z} - \frac{\partial P}{\partial z} \frac{\partial Q}{\partial \xi} \ . \tag{6.3}$$

One recognizes this at once as the standard Poisson bracket on a two-dimensional phase space with canonical coordinates $(z, \xi)$.

We must now take the classical limit of the generalized Adler map (3.9). The generalized Adler map can be rewritten as follows:

$$J^{(q)}(X) = \left[ \Lambda^{(q)}, X \right]_+ \circ \Lambda^{(q)} - \left[ \Lambda^{(q)}, (X \circ \Lambda^{(q)})_+ \right] \ , \tag{6.4}$$

which makes its classical limit obvious—namely,

$$J_{c\ell}^{(q)}(X) = [\![\Lambda^{(q)}, X]\!]_+ \Lambda^{(q)} - [\![\Lambda^{(q)}, (X\Lambda^{(q)})_+]\!] \ . \tag{6.5}$$

Expanding $J_{c\ell}^{(q)}(X)$ as

$$J_{c\ell}^{(q)}(X) = \sum_{i,j \geq 1} ((J_{c\ell}^{(q)})_{ij} \cdot x_j) \xi^{q-i} \ , \tag{6.6}$$

---

[3] We use $[\![,]\!]$ to the denote the Poisson bracket to avoid confusion with the Poisson bracket defining the $\mathsf{W}$-algebras.

we can read off the fundamental Poisson brackets. Notice that these consist of the terms in $J_{ij}^{(q)}$ with exactly one derivative. We can therefore read them off from (4.5) or else compute them from scratch using (6.5). Either way we obtain an expression which depends explicitly on $q$ and whose explicit form need not concern us here. As shown in [9] for the case $q = n \in \mathbf{N}$, this dependence in $q$ is fictitious and can be eliminated by a polynomial redefinition of variables. Just as in [9] the polynomial redefinition is easy to describe implicitly and we do so now.

Recall that two Poisson structures are said to be equivalent if there exists a polynomial redefinition of variables which renders the two structures proportional. We will show that for any nonzero $p, q$, the classical generalized Adler maps $J_{c\ell}^{(p)}$ and $J_{c\ell}^{(q)}$ are proportional.

Again let $\mathcal{M}_q$ denote the space of symbols of the form $\xi^q + \sum_{j \geq 1} u_j(z) \xi^{q-j}$. Then by the remark immediately following equation (2.23), the map $\varphi$ taking $\Lambda \mapsto \Lambda^{p/q}$ defines a map $\mathcal{M}_q \to \mathcal{M}_p$ with the property that the coefficients of $\Lambda^{p/q}$ are differential polynomials in the coefficients of $\Lambda$. The classical limit of this map is simply the $(p/q)^{\text{th}}$-product as commutative Laurent series, which for noninteger powers is defined as follows. Let $\Lambda = \xi^q + \sum_{j \geq 1} u_j(z) \xi^{q-j}$ and let us rewrite it as $\Lambda = (1 + \sum_{j \geq 1} u_j(z) \xi^{-j}) \xi^q$. It's $\alpha^{\text{th}}$-power is defined by

$$\Lambda^\alpha = \xi^{\alpha q} \exp\left(\alpha \log(1 + \sum_{j \geq 1} u_j \xi^{-j})\right) , \tag{6.7}$$

where both exp and log are defined by their power series around 0 and 1, respectively. From this definition it easily follows that if $\delta$ is any derivation,

$$\delta \Lambda^\alpha = \alpha \Lambda^{\alpha-1} \delta \Lambda . \tag{6.8}$$

The map $\varphi$ induces a map $\varphi_* : \mathcal{T}_q \to \mathcal{T}_p$ between tangent vectors and a dual map $\varphi^* : \mathcal{T}_p^* \to \mathcal{T}_q^*$ which are defined as follows. If $A \in \mathcal{T}_q$, then

$$\varphi_*(A) \equiv \left.\frac{d}{dt}(\Lambda + tA)^{p/q}\right|_{t=0} = \tfrac{p}{q} \Lambda^{p/q-1} A . \tag{6.9}$$

Similarly, if $X \in \mathcal{T}_p^*$, then $\varphi^*(X)$ is implicitly defined by

$$\langle \varphi^*(X), A \rangle = \langle X, \varphi_*(A) \rangle . \tag{6.10}$$

Using the fact that the classical limit of the bilinear form is simply given by the trace of the commutative product, we find

$$\varphi^*(X) = \tfrac{p}{q} \Lambda^{p/q-1} X . \tag{6.11}$$

With these two maps we can induce a Poisson structure $J_{c\ell}^?$ on $\mathcal{M}_p$ by completing the following commutative square:

$$\begin{array}{ccc} \mathcal{T}_q^* & \xrightarrow{J_{c\ell}^{(q)}} & \mathcal{T}_q \\ \uparrow \varphi^* & & \downarrow \varphi_* \\ \mathcal{T}_p^* & \xrightarrow{J_{c\ell}^?} & \mathcal{T}_p \end{array} \tag{6.12}$$

In other words, $J^?_{c\ell} : \mathcal{T}^*_p \to \mathcal{T}_p$ is given by $J^?_{c\ell} = \varphi_* \circ J^{(q)}_{c\ell} \circ \varphi^*$, where $\circ$ means here composition of maps. Explicitly, if $X \in \mathcal{T}^*_p$,

$$J^?_{c\ell}(X) = \left(\frac{p}{q}\right)^2 \Lambda^{p/q-1} J^{(q)}_{c\ell}\left(\Lambda^{p/q-1} X\right) , \qquad (6.13)$$

where $J^{(q)}_{c\ell}$ is given by (6.5) at $\Lambda^{(q)} = \Lambda$. Using repeatedly the fact that $\frac{p}{q}[\![\Lambda, \Lambda^{p/q-1} Z]\!] = \frac{p}{q}[\![\Lambda, Z]\!]\Lambda^{p/q-1} = [\![\Lambda^{p/q}, Z]\!]$ for any $Z$, we find that

$$\begin{aligned}
J^?_{c\ell}(X) &= \left(\frac{p}{q}\right)^2 \Lambda^{p/q-1} \left([\![\Lambda, \Lambda^{p/q-1} X]\!]_+ \Lambda - [\![\Lambda, (\Lambda^{p/q} X)_+]\!]\right) \\
&= \frac{p}{q} \left[\Lambda^{p/q} [\![\Lambda^{p/q}, X]\!]_+ - [\![\Lambda^{p/q}, (\Lambda^{p/q} X)_+]\!]\right] \\
&= \frac{p}{q} J^{(p)}_{c\ell}(X) \quad \text{evaluated at } \Lambda^{(p)} = \Lambda^{p/q},
\end{aligned} \qquad (6.14)$$

which proves the equivalence of $J^{(p)}_{c\ell}$ and $J^{(q)}_{c\ell}$ for any $p, q$.

In particular, this shows that all algebras in the one-parameter family have the same classical limit. In other words, the classical limit of $\mathsf{W}^{(q)}_{\mathrm{KP}}$ is $\mathsf{w}_{\mathrm{KP}}$ for all $q$ where this algebra is defined in [9]. Analogously, and after the $u_1(z) = 0$ reduction, the classical limit of $\widehat{\mathsf{W}}^{(q)}_\infty$ is independent of $q$ and yields a reduction of $\mathsf{w}_{\mathrm{KP}}$ denoted $\widehat{\mathsf{w}}_\infty$.

§7  Conclusions

Extending the Adler map to the space of pseudodifferential symbols of noninteger powers, we have constructed a one-parameter family (indexed by the highest power of the Lax-type operator) of hamiltonian structures for the KP hierarchy. These structures interpolate between the ones found by Radul, to which they reduce when the parameter is a positive integer. We have been so far unable, however, to promote them to a bihamiltonian pair by finding a suitable one-parameter family of coordinated brackets. Nevertheless, our results show that there are a lot more hamiltonian structures for the KP hierarchy than previously expected.

Under the identification of the fundamental Poisson brackets of hamiltonian structures of Lax-type hierarchies with W-algebras, this one-parameter family of hamiltonian structures gives rise to a one-parameter family $\mathsf{W}^{(q)}_{\mathrm{KP}}$ of W-algebras. This one-parameter family of W-algebras relates many known W-algebras via reductions and/or contractions and it is our hope that it plays an organizational role in the surveying of the topography of W-algebras of the $\mathsf{W}_\infty$-type. The relation between the many algebras connected by the one-parameter family constructed in this paper can be summarized by the following two commutative diagrams of W-algebras. The first diagram indicates the algebras obtained from $\mathsf{W}^{(q)}_{\mathrm{KP}}$ at special values of $q$:

$$
\begin{array}{ccc}
& & \mathsf{W}_{\mathrm{KP}} \xrightarrow{\text{reduction}} \widehat{\mathsf{W}}_\infty \\
& & \uparrow \text{evaluation } q=1 \quad\quad \uparrow \text{evaluation } q=1 \\
& & \mathsf{W}_{\mathrm{KP}}^{(q)} \xrightarrow{\text{reduction}} \widehat{\mathsf{W}}_\infty^{(q)} \\
& & \downarrow \text{evaluation } q=n \quad\quad \downarrow \text{evaluation } q=n \\
\mathsf{GD}_n^{(1)} \times \mathsf{W}_{1+\infty}^{c=0} & \xleftarrow{\text{contraction}} & \mathsf{W}_{\mathrm{KP}}^{(n)} \xrightarrow{\text{reduction}} \widehat{\mathsf{W}}_\infty^{(n)} \\
\downarrow \text{reduction} & & \downarrow \text{reduction} \quad\quad \downarrow \text{reduction} \\
\mathsf{GD}_n^{(1)} & \xleftarrow{\text{contraction}} & \mathsf{GD}_n \xrightarrow{\text{reduction}} \mathsf{W}_n
\end{array}
$$

where the horizontal arrows labelled "reduction" correspond to the reduction induced by setting the field of lowest weight to zero. The contraction from $\mathsf{W}_{\mathrm{KP}}^{(n)}$ is the first Dickey–Radul hamiltonian structure, which breaks up as two commuting subalgebras: one isomorphic to a centerless $\mathsf{W}_{1+\infty}$ and the other being the first Gel'fand–Dickey bracket. Of course, this diagram can be further extended by considering reductions of $\mathsf{GD}_n$ induced from imposing definite symmetry conditions on the Lax operator. These W-algebras are associated to the $B$ and $C$ series of simple Lie algebras in the same way that $\mathsf{W}_n$ is associated to the $A$ series [**23**].

The second diagram indicates those algebras reached via contractions of $\mathsf{W}_{\mathrm{KP}}^{(q)}$ and its reductions:

$$
\begin{array}{ccccccc}
\mathsf{W}_{\infty-N} & \xleftarrow{\text{truncation}} & \mathsf{W}_\infty & \xleftarrow{\text{truncation}} & \mathsf{W}_{1+\infty} & \xrightarrow{\text{reduction}} & \mathsf{W}_\infty^\# \\
& \nwarrow \text{contraction } q\to N & & & \uparrow \text{contraction } q\to 0 & & \uparrow \text{contraction } q\to 0 \\
& \widehat{\mathsf{W}}_{\infty-N}^{(q)} & \xleftarrow{\text{reduction}} & & \mathsf{W}_{\mathrm{KP}}^{(q)} & \xrightarrow{\text{reduction}} & \widehat{\mathsf{W}}_\infty^{(q)} \\
& & & & & & \downarrow \text{contraction } q\to 1 \\
& & & & & & \mathsf{W}_\infty
\end{array}
$$

The algebra $\mathsf{W}_\infty^\#$ is a new genuinely nonlinear algebra which extends the Virasoro algebra with generators of spins $\geq 3$. It is obtained as a reduction from $\mathsf{W}_{1+\infty}$ by setting the spin-one generator to zero. The reduction to $\widehat{\mathsf{W}}_{\infty-N}^{(q)}$ is obtained by setting the generators with spins $\leq N$ equal to zero. This reduction yields in general a nonlocal Poisson algebra, but upon contraction the nonlocal terms, as do the nonlinear ones, vanish.

Under classical limits we found that the $q$-dependence disappears and all classical algebras are isomorphic. This situation can again be summarized by the following commutative diagram

$$
\begin{array}{ccc}
\mathsf{W}_{\mathrm{KP}}^{(q)} & \xrightarrow{\text{reduction}} & \widehat{\mathsf{W}}_\infty^{(q)} \\
\downarrow \text{classical limit} & & \downarrow \text{classical limit} \\
\mathsf{w}_{\mathrm{KP}} & \xrightarrow{\text{reduction}} & \widehat{\mathsf{w}}_\infty
\end{array}
$$

where the horizontal arrows have the same meaning as before.

While in the process of typing this paper, a paper appeared [24] containing a deformation of $\widehat{\mathsf{W}}_\infty$ based on the two-boson realization of [12]. Nevertheless, the deformation in [24] is such that for all nonzero values of the parameter the algebra remains isomorphic to $\mathsf{W}_{\mathrm{KP}}$ and for the parameter tending to zero, the algebra contracts to a centerless $\mathsf{W}_\infty$. Therefore, the results described in this paper provide the first nontrivial deformation of $\mathsf{W}_{\mathrm{KP}}$ (or $\widehat{\mathsf{W}}_\infty$).

To conclude, we would like to stress that using our results one can find a continuous link between the nonlinear $\mathsf{W}_{\mathrm{KP}}$ algebra and the $w_\infty$ algebra (obtained by further contracting $\mathsf{W}_\infty$). As mentioned in the introduction, the former algebra—or rather a quantization thereof—is the chiral algebra of the noncompact coset model describing the black hole solution of Witten; whereas the latter algebra is known to be relevant for the dynamics of the $c = 1$ matrix model and shows up as well in the continuum. If these two models are supposed to represent different phases of two-dimensional gravity coupled to $c = 1$ conformal matter, it raises questions on the nature of the corresponding phase transition. It is plausible that further study in this direction will unveil the relevance of the mentioned infinite-dimensional algebras as underlying dynamical principles of the corresponding models. We think that the algebraic link that we have set up in this paper will prove useful in this context.

# ACKNOWLEDGEMENTS

E. R. would like to express his gratitude to the Departamento de Física de Partículas Elementales de la Universidad de Santiago de Compostela for its hospitality and support during this collaboration. J. M. F. would like to do the same to the Instituut voor Theoretische Fysica of the Universiteit Leuven for the same reasons. We would also like to extend our thanks to S. Stanciu who helped us check some calculations.